\documentclass[twocolumn,prl,aps,showpacs,superscriptaddress]{revtex4}
\usepackage{graphicx}% Include figure files
\usepackage{amsmath} 
\usepackage{amsthm} 
\usepackage{graphicx} 
\usepackage{float}
\usepackage{amssymb}

\begin{document}

\title{On the effect of localized and delocalized quasiparticles on a nonlinear conductivity of superconductors}

\author{A. Gurevich}
\affiliation{Department of Physics and Center for Accelerator Science, Old Dominion University, Norfolk VA 23510}

\date{\today}%

\begin{abstract}
Recently Deyo et al. [\prb {\bf 106}, 104502 (2022)] suggested that the surface resistance $R_s(H_0)$ of a superconductor 
under strong electromagnetic field can be affected by field-induced quasiparticle bound states at the surface. 
This Comment shows that the existence of such bound states and their contribution to $R_s$ are not    
substantiated and the phenomenological models of Ref. \cite{corn} give incorrect field and temperature dependencies of $R_s(T,H_0)$.

\end{abstract}


\maketitle

Deyo {\it et al.} analyzed the surface states in a clean superconductor, using the Bogoliubov-de-Gennes (BdG) equations 
for a film of thickness $L=5\lambda$ in a magnetic field $H(t)=H_0\sin\omega t$ applied along the $x$-axis:
\begin{gather}
\!\!\!\!\frac{\hbar^2}{2m}\left[k_x^2+\left(k_y-\frac{eA}{\hbar c}\right)^2\!\!-\frac{d^2}{dz^2}\right]\!u
+\Delta v=(E+E_F)u,
\label{b1} \\
\!\!\!\!\frac{\hbar^2}{2m}\left[k_x^2+\left(k_y+\frac{eA}{\hbar c}\right)^2\!\!-\frac{d^2}{dz^2}\right]\!v
-\Delta u=(E_F-E)v,
\label{b2}\\
u(0)=v(0)=0,\qquad u(L)=v(L)=0.
\label{bc}
\end{gather}
Here $[u({\bf r}), v({\bf r})]=[u(z),v(z)]e^{ik_xx+ik_yy}$, $A(z,t)=H_0\lambda e^{-z/\lambda}\sin\omega t$, $\lambda$ is the London penetration depth, and the energy $E$ is counted from the Fermi surface. 

\begin{figure}
\centering
\includegraphics[width=8cm, trim={0mm 0mm 0mm 0mm},clip]{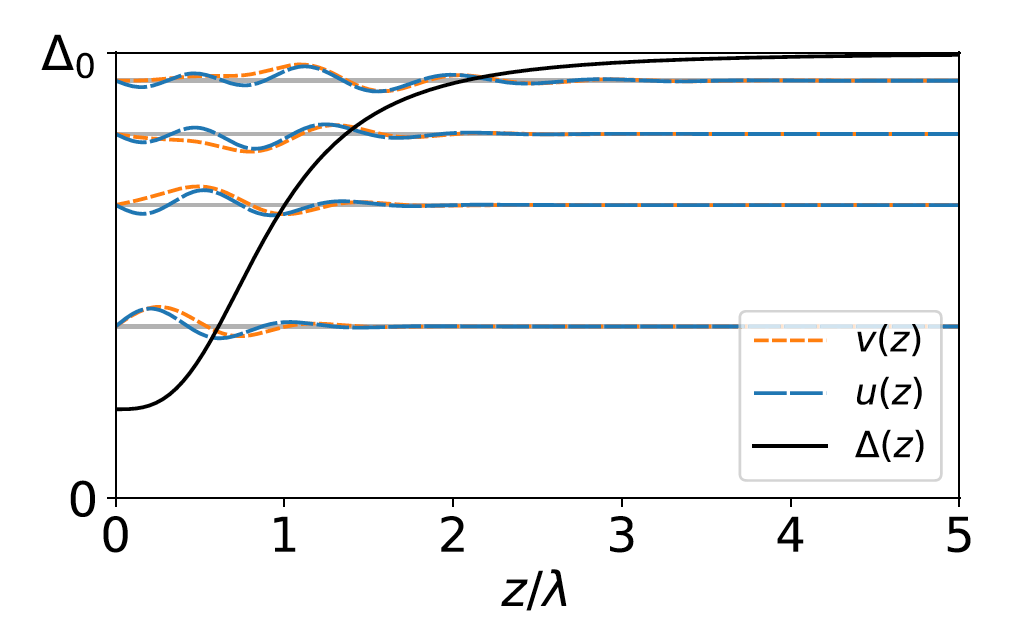}
   \caption{
The BdG quasiparticle bound states for $k_x = k_F$, $k_y = 0$,
at a field of 40 MV/m. For each state the wave functions $u$ and $v$ are
plotted with a vertical shift according to their energy. The situation
is somewhat analogous to a particle in a potential $\Delta(z)$ with a hard
wall on the left. Reproduced from Ref. \cite{corn}.
   }\label{fig1}
\end{figure}

Fig. {\ref{fig1}} shows the bound states in a deep well in $\Delta(z)$ for quasiparticles moving parallel to the surface with velocities $v_x=v_F$ obtained in Ref. \cite{corn}. Here the well in $\Delta(z)$ is caused by the Meissner currents but not the boundary conditions (\ref{bc}) as it only 
occurs at the film side exposed to the magnetic field.  The strong suppression of $\Delta(z)$ at the surface shown in Fig. \ref{fig1} appears inconsistent with the magnitude of current pairbreaking \cite{tinkh}. This can be seen from the Ginzburg-Landau (GL) equation,
\begin{equation}
\xi^2\psi''+\bigl[1-\left(2\pi\xi\lambda H_0/\phi_0\right)^2e^{-2z/\lambda}\bigr]\psi-\psi^3,
\label{gl}
\end{equation}  
where $\psi(z)=\Delta(z)/\Delta_0$, $\Delta_0$ is the order parameter at zero field, $\xi$ is the coherence length, and $\phi_0$ is the flux quantum. The term $\propto H_0^2$ in Eq. (\ref{gl}) causes a weak suppression in $\psi(z)=1+\delta\psi(z)$, where the field-induced correction $\delta\psi(z)$ satisfies the linearized GL equation: 
\begin{equation}
\xi^2\delta\psi''-2\delta\psi=(H_0/\sqrt{2}H_c)^2e^{-2z/\lambda},
\label{glg}
\end{equation} 
where $H_c=\phi_0/2^{3/2}\xi\lambda$ is the GL thermodynamic critical field. The solution with $\Delta'(0)=0$ like in Fig 1 is:
\begin{equation}
\frac{\Delta(z)}{\Delta_0}=1+\frac{(H_0/2H_{c})^{2}}{2(\xi/\lambda)^{2}-1}\biggl[e^{-2x/\lambda}-\frac{\xi\sqrt{2}}{\lambda}e^{-\sqrt{2}x/\xi}\biggr].
\label{gla}
\end{equation}
If $\kappa=\lambda/\xi\gg 1$ the well in $\Delta(z)$ at $H_0=H_{c1}$ is very shallow, $\delta\psi\sim -\kappa^{-2}$. Likewise, $\delta\psi\sim -2^{-5/2}\kappa$  at $\kappa\ll 1$ and $H_0=H_c$. For a clean Nb with $\xi=\lambda$, Eq. (\ref{gla}) also gives a very shallow well $\Delta(0)/\Delta_0=0.994-0.974$ in the field region $H_0=(0.25-0.5)H_c$, where the decrease of $R_s(H_0)$ with $H_0$ in Nb cavities has been observed \cite{raise}. For $H_0\approx H_c^{Nb}=200$ mT, Eq. (\ref{gla}) gives $\Delta(0)/\Delta_0\approx 0.9$, much different from $\Delta(0)/\Delta_0\approx 0.25$ in Fig. \ref{fig1}.  At $T\ll T_c$ the well in $\Delta(z)$  in the clean limit is even shallower than the above GL estimates because a uniform current has no effect on $\Delta$ at $T=0$ and the superfluid velocities $v_s<v_c=\Delta/p_F$  \cite{maki}. However, to obtain this result, the calculation of $\Delta(z)$ at $T\ll T_c$ requires not only summations over $k_x$ and $k_y$ but also taking into account lots of ($\sim Lk_F\sim 10^3$ for Nb) rapidly oscillating eigenfunctions $u_n(z)$ and $v_n(z)$ with periods $l({\bf v}_F)\sim \hbar/mv_z \ll \lambda$ and  $|E|<2\hbar\omega_D$, where $v_z$ is the $z$ component of the Fermi velocity ${\bf v}_F$.  There is no information in Ref. \cite{corn} about how the results shown in Fig. \ref{fig1} were produced and what eigenstates were taken into account in the self-consistency equation for $\Delta(z)$. If only $u(z)$ and $v(z)$ varying slowly over $\lambda_F=2\pi/k_F$ were taken into account, while lots of itinerant states with $v_z\sim v_F$ were disregarded, as was briefly mentioned in Ref. \cite{corn}, the bound states shown in Fig. \ref{fig1} are not real.   

Equations (\ref{b1})-(\ref{b2}) are ill-suited for the numerical calculations of itinerant states $u({\bf r})=e^{i{\bf k_Fr}}\alpha({\bf r})$ and $v({\bf r})=e^{i{\bf k_Fr}}\beta({\bf r})$, where  $\alpha({\bf r})$ and $\beta({\bf r})$ vary slowly over $\lambda_F$.  In this quasi-classical Andreev approximation the BdG equations become:
\begin{gather}
-i\hbar v_{z}\alpha'+v_{y}p_s\alpha+\Delta\beta=E_-\alpha
\label{a3} \\
i\hbar v_{z}\beta'+v_{y}p_s\beta+\Delta^{*}\alpha=E_+\beta
\label{a4}
\end{gather}
Here $E_{\pm}=E\pm \hbar(v_{x}s_{x}+v_{y}s_{y})$, ${\bf s}={\bf k}-{\bf k}_F$, and ${\bf p}_s=m\hbar {\bf v}_s$ defines the Meissner current density ${\bf J}=en_s{\bf v}_s$.
If $p_s$ is independent of $z$, the solutions $(\alpha,\beta)\propto e^{is_{z}z}$ give the well-known anisotropic  energy 
spectrum  \cite{bardeen,maki,tinkh}:
\begin{equation}
E({\bf k})=p_sv_{z}\pm\sqrt{\xi_{{\bf k}}^{2}+\Delta^{2}},
\label{eni}
\end{equation}
where $\xi_{\bf k}=\hbar(({\bf k-k}_F)\cdot{\bf v}_F$. As follows from Eq. (\ref{eni}), the energy gap $E_g({\bf v}_F)=\Delta_0-{\bf v}_F{\bf p}_s$ depends on the direction of ${\bf v}_F$ relative to ${\bf J}$, the minimum gap $E_g(t)=\Delta_0-|p_s(t)|v_F$ is smaller than $\Delta_0$ \cite{bardeen}. 
Inserting Eq. (\ref{eni}) in the self-consistency equation for $\Delta$ and integrating over the orientations of ${\bf v}_F$ at $T=0$
gives $\Delta$ independent of $p_s$ if $p_s<\Delta/v_F$ \cite{maki}. This also follows from the Eilenberger equations \cite{eli}. If $\alpha(z)$ and $\beta(z)$ vary slowly over $\lambda_F$ in the case of $p_s(z)=(e\lambda H_0/c)e^{-z/\lambda}$, the eigenstates of Eqs. (\ref{a3})-(\ref{a4}) 
are in a narrow energy belt around the Fermi surface where the BCS pairing occurs.

The dependencies of the surface energy levels on $(k_x,k_y)$ and $H_0$ were not investigated in Ref. \cite{corn}, where Fig. 1 was presented  
as the only evidence for such states. 
Since the well in $\Delta(z)$ calculated self-consistently is likely much shallower than $\Delta(z)$ shown in Fig. \ref{fig1},  
the existence of these bound states is questionable. Even if such weakly-bound stakes might occur at $H_0\simeq H_c$, they would get hybridized with the itinerant states due to impurity scattering or surface topographic defects because the energy gap $E_g(H_0)=\Delta_0-v_x|p_s(t)|$ for itinerant quasiparticles is lower than the bottom of a shallow well in $\Delta(z)$. For instance, in the clean limit at $\kappa\gg 1$ and $T=0$ the gap $E_g(H_0)=\Delta_0(1-H_0/H_g)$ decreases with $H_0$ and vanishes at $H_g=(2/3)^{1/2}H_c$, while $\Delta$ remains constant \cite{eli}. In any case, the density of itinerant quasiparticles with $E_g<E<\Delta_0$ and different orientations of ${\bf v}$ is much greater than the  density of bound states (if any) in which ${\bf v}$ is parallel to the  surface, so the contribution of bound states to rf losses is negligible.  

To evaluate the surface resistance $R_s(H_0)$ Deyo {\it et al.} used a "three-liquid" model in which the dissipative part of conductivity was taken in the Drude-like form,
\begin{equation}
\sigma_1=(n_c\tau_c+n_b\tau_b)e^2/m,
\label{drud}
\end{equation}
where $n_b$ and $n_c$ denote the densities of bound and continuum
states, and $\tau_b$ and $\tau_c$ are their respective relaxation times. Here $n_c$ and $n_b$ were evaluated by summing up the quasiparticle states but how it was done and what states were taken into account were not specified in Ref. \cite{corn}. Moreover, it was stated that $E <\Delta_0$ belong to bond states and $E>\Delta_0$ to itinerant states, which contradicts the fact that the energies $E_g(H_0)<E<\Delta_0$ actually   
belong to continuum states. Unlike the putative surface bound states, the itinerant quasiparticles with $E_g(H)<E<\Delta_0$ do contribute significantly to $R_s$.  The phenomenological Eq. (\ref{drud}), which is not obtained from the theory of electromagnetic response of superconductors \cite{mb,tinkh,kopnin}, gives incorrect temperature and field dependencies of $\sigma_1(T,H_0)$. For instance, at low fields Eq. (\ref{drud}) yields $\sigma_1\propto n_c\propto T^{1/2}e^{-\Delta/T}$ inconsistent with the Mattis-Bardeen $\sigma_1\propto T^{-1}\ln(T/\hbar\omega)e^{-\Delta/T}$ which has a logarithmic singularity at $\omega\to 0$.  

The problems with the field dependence of $\sigma_1(H_0)$ given by Eq. (\ref{drud}) are apparent if $\tau_b=\tau_c$ and the sum $n_c+n_b$ combines to the total density of thermally-activated quasiparticles $n_{qp}(H_0)$. Since $n_{qp}(H_0)$ increases with $H_0$  \cite{agp}, Deyo {\it et al.} concluded that itinerant quasiparticles cannot produce the decrease of $R_s(H_0)$ with $H_0$ observed on Nb cavities.  However, this just reflects the inadequacy of Eq. (\ref{drud}) because $\sigma_1(H_0)$ decreasing with $H_0$ has been derived from a microscopic theory \cite{ag,kg}. To obtain the decrease of $R_s(H_0)$ with $H_0$ in their model Deyo {\it et al.} not only ascribed $E>\Delta_0$ and $E<\Delta_0$ to continuum and bound states, but also assumed that these states have different relaxation times $\tau_c$ and $\tau_b$. As a result,
$\sigma_1(H_0)$ can decrease with $H_0$ if $\tau_b\ll\tau_c$  because the density of quasiparticles with $E>\Delta_0$ and large $\tau_c$ decreases while the  density of quasiparticles with $E<\Delta_0$ and small $\tau_b$ increases as $H_0$ increases.  

Assigning unequal $\tau_b$ and $\tau_c$ for itinerant states at $E_g<E<\Delta_0$ and $E>\Delta_0$, and using various time constants including inelastic electron-phonon (e-p) collision times $\tau_{ep}$ \cite{kaplan} as adjustable parameters in the phenomenological Eq. (\ref{drud}) in Ref. \cite{corn} has no theoretical basis.  Generally, $\sigma_1$ is determined by the rf power absorbed by quasiparticles and is independent of $\tau_{ep}$ at low $\omega$ and $H_0$, whereas $\sigma_1$ at $H_0\ll H_c$ in the clean limit is independent of any elastic or inelastic relaxation times \cite{mb}. The e-p collisions do control the energy transfer from quasiparticles to phonons, producing a nonequilibrium correction $\delta f(E,t)$ to the quasiparticle distribution function which becomes significant at very low $T$ or high $\omega$ and $H_0$. Yet in the range of the parameters, where $\delta f(E,t)$ is much smaller than the equilibrium $f_0(E)=\left(e^{E/T}+1\right)^{-1}$, the e-p collisions do not affect $R_s$ determined mostly by elastic impurity scattering and the effects of current on the density of states and the coherence factors. These key effects are not properly taken into account in Eq. (\ref{drud}). 

A microscopic theory of nonlinear surface resistance based on the Keldysh technique for nonequilibrium superconductors in the dirty limit at $\lambda\gg\xi$ \cite{kopnin} was developed in Refs. \cite{ag,kg}.  The results of Refs. \cite{ag,kg} were misinterpreted by Deyo {\it et al.} who claimed that they rely crucially on the "notion of  maintaining $E_2=E_1+\hbar\omega$", "approximation to determine the nonequilibrium quasiparticle distribution function" or "The smeared density of states",  implying that the spatial variation of screening current and temporal oscillations of the superconducting gap, density of states and coherence factors were disregarded in Refs. \cite{ag,kg}.  Neither of these assertions is relevant, as one can see from the formulas for $R_s$ in the case of {\it equilibrium} $f_0(E)$ \cite{ag,kg}:   
\begin{gather}
\!\!R_s=R_1\!\!\int_0^{\pi/\omega}\!\!\!\!dt\!\int_0^{b_0}\!db\!\int_{E_g(t)}^\infty\frac{M[E, s(t)]dE}{\cosh(\frac{E}{2T})\cosh[\frac{(E+\hbar\omega)}{2T}]},
\label{rs}\\
M=\mbox{Re}G[\hbar\omega+E, s(t)]\mbox{Re}G[E, s(t)] +
\nonumber \\
\mbox{Re}F[\hbar\omega+E, s(t)]\mbox{Re}F[E, s(t)],
\label{M}\\
s=b\sin^2 \omega t,\qquad E_g(t)=[\Delta^{2/3}(t)-s^{2/3}(t)]^{3/2},
\label{eg}
\end{gather}
where $R_1=\mu_0^2\omega^2\lambda^3\sinh(\hbar\omega/2T)/2\pi\hbar\rho_n$, $E_g(t)$ is the quasiparticle gap, $b=b_0e^{-2z/\lambda}$, $b_0=(H_0/2H_c)^2\Delta_0$, $H_c=\phi_0/2^{3/2}\xi\lambda$, $\Delta(t)=\Delta_0-\pi s(t)/4$, and the Green functions $G^R(E,t)$ and $F^R(E,t)$ were obtained by solving the time-dependent Usadel equation at $\hbar\omega\ll \Delta$ \cite{ag}. The energy gap $E_g(z,t)$ is always lower than the bottom of the well in $\Delta(z,t)$. Here $E_g(t)$, $\Delta(t)$ and the spectral function $M(E,t)$ which accounts for the density of states and the coherence factors, oscillate under rf field so no "smeared density of states" is involved here. The integration over $b$ accounts for the spatial change of $J(z,t)$ with the distance from the surface \cite{ag}. The arguments $E$ and $E+\hbar\omega$ in $M(E,t)$ have nothing to do with constant quasi-energies but result from a straightforward calculation of the rf power $R_sH_0^2/2=\int_0^\infty \langle J(z,t)\epsilon(z,t)\rangle dz$, where $\langle...\rangle$ denotes time averaging over the rf period,  
$\epsilon(z,t)=(H_0\lambda\omega/c) e^{-z/\lambda}\cos\omega t$ is the rf electric field and the superconducting current $J(z,t)$ \cite{kopnin} was calculated using analytical solutions for $G^R(E,t)$ and $F^R(E,t)$ given in Refs. \cite{ag,kg}. Here $E$ just labels the quasiparticle states in the time-dependent spectral function $M(E,t)$.   

\begin{figure}[ht]
\centering
\includegraphics[width=8.3cm, trim={20mm 75mm 20mm 55mm},clip]{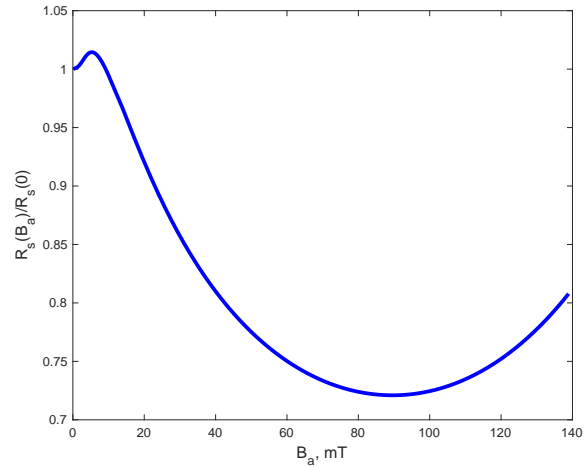}
   \caption{
The surface resistance calculated from Eqs. (\ref{rs})-(\ref{eg}) for Nb at $\Delta_0 =18$ K, $T=2K$ and 1.3 GHz.
   }\label{fig2}
\end{figure}

Shown in Fig. \ref{fig2} is an example of $R_s(H_0)$ calculated from Eqs. (\ref{rs})-(\ref{eg}) which captures the nonmonotonic field dependence of $R_s(H_0)$ observed on Nb cavities \cite{raise,ag,kg}. Such generic behavior of $R_s(H_0)$ occurs even in the case of equilibrium $f_0(E)$ of itinerant quasiparticles driven by a low-frequency rf current at $\omega\tau_{ep}\lesssim 1$, although a frozen density of nonequilibrium quasiparticles at $\omega\tau_{ep}\gg 1$ may cause even stronger decrease of $R_s(H_0)$ with $H_0$ \cite{ag}.  Developing a microscopic theory of $R_s(H_0)$ of a nonequlibrium superconductor at low $T$ and high $\omega$ is a challenging outstanding problem.

\vspace{1mm}

In conclusion, the surface bound states shown in Fig. 1 and their contribution to the nonlinear electromagnetic response are questionable. The  phenomenological Eq. (\ref{drud}) cannot serve as a theoretical basis for the calculations of $R_s(H_0)$, particularly the field-induced reduction of $R_s(H_0)$ which has been observed on Nb cavities.    

This work was supported by DOE under Grant DE-SC0010081-020.

\end{document}